\documentclass[%
 reprint,
groupedaddress,
 amsmath,amssymb,
]{revtex4-1}

\usepackage{dsfont}
\usepackage{graphicx}
\usepackage{hyperref}

\newcommand{\im}{\mathrm{i}}
\newcommand{\idop}{\mathds{1}}

\newcommand{\argmax}[1]{\underset{#1}{\text{argmax}}}

\begin{document}
\title{Effective time reversal and echo dynamics in the transverse field Ising model}

\author{Markus Schmitt}
\email{markus.schmitt@theorie.physik.uni-goettingen.de}
\author{Stefan Kehrein}
\affiliation{%
 Institute for Theoretical Physics, 
	Georg-August-Universit\"at G\"ottingen 
	- Friedrich-Hund-Platz 1, G\"ottingen 37077, Germany
}
\date{\today}


\begin{abstract}
The question of thermalisation in closed quantum many-body systems has received a lot of attention in 
the past few years. An intimately related question is whether a closed quantum system shows 
irreversible dynamics. However, irreversibility and what we actually mean by this in a quantum 
many-body system with unitary dynamics has been explored very little. In this work we investigate the 
dynamics of the Ising model in a transverse magnetic field involving an imperfect effective time reversal. 
We propose a definition of irreversibility based on the echo peak decay of observables. 
Inducing the effective time reversal by different protocols we find algebraic decay of the
echo peak heights or an ever persisting echo peak indicating that the dynamics in this model is well
reversible.
\end{abstract}

\maketitle

\section{Introduction}
During the last decades enormous advances in the experimental realisation of
highly controllable quantum simulators \cite{Greiner2002,Kinoshita2006,Bloch2012, Georgescu2014} 
have triggered a lot of activity in 
theoretically investigating the
out of equilibrium dynamics of quantum many-body systems. In particular the
equilibration of closed many-body systems and the process of thermalisation as
fundamental questions of quantum statistical mechanics aroused a lot of interest 
\cite{Manmana2007,Rigol2007,Rigol2008,Moeckel2008,Reimann2008,Calabrese2011,
Fagotti2013,Caux2013,Polkovnikov2011,Eisert2015}.
Nevertheless, albeit being intimately related to thermalisation the question of 
irreversibility in quantum many-body systems has to date hardly been addressed.

In the context of classical systems this question was already discussed during the
development of thermodynamics. Regarding Boltzmann's H-theorem \cite{Boltzmann1872} Loschmidt
pointed out that in his derivation of the Second Law Boltzmann had obviously broken
the time reversal invariance of the underlying microscopic laws of motion\cite{Loschmidt1876}.
Specifically, he argued that if one performs an effective time
reversal on a classical gas by inverting the velocities of all particles 
at some point in time the system must
necessarily return to its initial state after twice that time. With this example at hand the
emergence of irreversibility in classical systems is nowadays easily understood: A system 
with sufficiently many degrees of freedom will generically exhibit chaotic dynamics and
therefore any time reversal operation will be practically infeasible due to the exponential
sensitivity of the dynamics to inevitable errors.
This is also a way to understand the loss of information about the initial state during
the time evolution, which is essential for thermalisation. In a chaotic many-body system with irreversible
dynamics there is no realisable protocol that would allow to return it to the initial state.

Referring to the knowledge about classical irreversibility Peres suggested to study the 
Loschmidt echo
\begin{align}
\mathcal L(\tau)=|\langle\psi_0|e^{\im (H+\epsilon V)\tau}e^{-\im H\tau}|\psi_0\rangle|^2
\end{align}
in order to quantify irreversibility of quantum systems \cite{Peres1984}. 
The Loschmidt echo is the overlap of
the initial state with the forward and backward time evolved state when including a small 
deviation $\epsilon V$ in the time evolution operator of the backwards evolution.
As such it quantifies how well the initial state is resembled after an imperfect effective time reversal.
The Loschmidt echo turned out to be a very interesting measure when studying systems with
few degrees of freedom, exhibiting a variety of possible decay characteristics
\cite{Gorin2006,Jacquod2009}.

However, in generic quantum many-body systems the Loschmidt echo is not a measurable quantity.
If the prerequisite of the Eigenstate Thermalisation Hypothesis (ETH) \cite{Deutsch1991, Srednicki1994, Srednicki1996} pertains, which all 
numerical evidence indicates \cite{Rigol2008,Steinigeweg2014,Beugeling2014}, 
then expectation values of local observables 
$O_E=\langle E|\hat O|E\rangle$ are smooth functions of the eigenstate energy $E$. This means that 
even orthogonal states cannot necessarily be distinguished experimentally.
This argument carries over to integrable systems when the observable expectation value is 
considered as a function of all integrals of motion instead of only the energy \cite{Cassidy2011}.
Therefore a definition of irreversibility with respect to the Loschmidt-echo cannot meaningfully
differentiate between reversible and irreversible dynamics in many-body systems.
It should also be noted that generally
the Loschmidt echo is of large deviation form, $\mathcal L(\tau)\sim e^{-Nl(\tau)}$ 
with some rate function $l(\tau)$, i.e. it is exponentially suppressed with increasing system size $N$. 

In our work, when addressing the question of irreversibility in many-body systems we focus on 
observable
echoes that are produced under imperfect effective time reversal, i.e.
\begin{align}
	\langle O\rangle_\tau&=\langle\psi(\tau)|\hat O|\psi(\tau)\rangle\ ,\nonumber\\
	|\psi(\tau)\rangle&=e^{\im (H+\epsilon V)\tau}e^{-\im H\tau}|\psi_0\rangle\ .
\end{align}
We propose a definition of irreversibility based on the decay of the echo peak as the waiting time
$\tau$ is increased. With respect to that we consider the dynamics of systems exhibiting an 
algebraic decay reversible, whereas systems with exponentially or faster than exponentially decaying
echo peaks are irreversible.

Obviously, echoes in the expectation values of observables will depend on the choice of the observables.
Thus, the conclusions that can be drawn regarding the irreversibility of the dynamics will have to
be decided on a case by case basis. 
However, to the best of the current knowledge fundamental issues of thermalisation, in particular
the description of stationary expectation values in unitarily evolved pure states after long times by
thermal density matrices, can likewise only be understood for specific classes of observables
\cite{Polkovnikov2011,Eisert2015}.

Recently, an alternative definition for chaos in quantum systems was put forward, which is based on the 
behaviour of
out-of-time-order (OTO) correlators of the form $\langle W(t)V(0)W(t)V(0)\rangle$. These OTO
correlators probe a system's sensitivity to small perturbations \cite{Maldacena2015}. 
Moreover, they are closely related
to the phenomenon of scrambling, i.e. the complete delocalisation of initially local information under
time evolution \cite{Hosur2016}. The relation between both definitions should be investigated
systematically in future work.

An important experimental application of effective time reversal are NMR experiments. The
dynamics of non-interacting spins can be reverted by the Hahn echo technique \cite{Hahn1980} or 
by the application of more sophisticated pulse sequences \cite{Haeberlen1976,Uhrig2007}. 
Moreover, it is possible to realise effective time reversal in certain dipolar coupled spin systems
by the so called magic echo technique \cite{Schneider1969,Rhim1971,Hafner1996}. 
Particularly notable are various experimental and theoretical works on the refocussing of a local
excitation by effective time reversal in NMR setups 
\cite{Zhang1992,Pastawski1995,Levstein1998,Zangara2015}. 
Besides that we expect that effective time reversal can be realised in quantum simulators 
\cite{Bloch2012, Georgescu2014}; and recently there
were proposals for effective time reversal by periodic driving \cite{Mentink2015} or
by spin flips in cold atom setups with spin-orbit coupling \cite{Engl2014}.

Results for the echo dynamics in many-body systems might also be interesting from other points of
view. For example, there are proposals for the identification of many-body localised phases using
spin echoes \cite{Serbyn2014} or for the certification of quantum simulators using effective time 
reversal \cite{Wiebe2014}.


In this letter we report results for effective time reversal in the transverse field Ising model (TFIM).
This simple model Hamiltonian is diagonal in terms of fermionic degrees of freedom and all
quantities of interest can be computed analytically in the thermodynamic limit. Thus, it has well known 
properties and, in particular, the stationary state
it approaches in the long time limit is well understood \cite{Calabrese2011,Calabrese2012a}. 
As such the TFIM is ideally suited as a starting point to study irreversibility theoretically from the
aforementioned point of view.
On top of this, the TFIM has been realised experimentally in circuit QED \cite{Viehmann2013}.

\section{Dynamics in the transverse field Ising model}
The Ising model in a transverse magnetic field is defined by the Hamiltonian
\begin{align}
	H(h)=-J\sum_{i=1}^NS_i^zS_{i+1}^z+h\sum_{i=1}^NS_i^x\ ,\label{eq:tfim_ham}
\end{align}
where $S_i^{x/z}$ denotes the Pauli spin operators acting on lattice site $i$, $N$ the number of lattice
sites, and $h$ the magnetic field strength \cite{Pfeuty1970}. For our purposes we consider periodic
boundary conditions.
A Jordan-Wigner transform allows to map this spin 
Hamiltonian to a quadratic Hamiltonian in momentum space
\begin{align}
	H(g)=J\sum_{k>0}
	\begin{pmatrix}c_k^\dagger&c_{-k}\end{pmatrix}
	\begin{pmatrix}d_k^z(g)&-\im d_k^y\\\im d_k^y&-d_k^z(g)\end{pmatrix}
	\begin{pmatrix}c_k\\c_{-k}^\dagger\end{pmatrix}
\end{align} 
with fermionic operators $c_k^\dagger,c_k$ and coefficient functions
$d_k^y(g)=\sin(k)/2$ and $d_k^z(g)=g-\cos(k)/2$, where $g=h/J$.
The Bogoliubov rotation
\begin{align}
	\begin{pmatrix}
		\lambda_k\\\lambda_{-k}^\dagger
	\end{pmatrix}
	=R^x(\theta_k^g)
	\begin{pmatrix}
		c_k\\ c_{-k}^\dagger
	\end{pmatrix}
\end{align}
with Bogoliubov angle $\theta_k^g=\arctan\left(d_k^y/d_k^z(g)\right)$
diagonalises the Hamiltonian yielding
\begin{align}
	H(g)=\sum_{k>0}\epsilon_{k}^g\lambda_k^\dagger\lambda_k
\end{align}
with energy spectrum $\epsilon_k^g=J\sqrt{d_k^y(g)^2+d_k^z(g)^2}$.
The gap closing point at $g=1/2$ indicates the quantum phase transition between paramagnet
and ferromagnet.

Above a family of unitary matrices,
\begin{align}
	R^\alpha(\phi)=\idop\cos\frac{\phi}{2}+\im\sigma^\alpha\sin\frac{\phi}{2}\ ,\quad \alpha\in\{x,y,z\}\ ,
	\label{eq:unitaries}
\end{align}
with the Pauli matrices $\sigma^\alpha$ was introduced for later convenience.

After mapping the spin degrees of freedom to free fermions expectation values of many observables
are -- thanks to Wick's theorem --
given in terms of block Toeplitz (correlation) matrices $\Gamma_{ij}\equiv\Gamma_{(i-j)}$, where
\begin{align}
	\Gamma_l=\begin{pmatrix}
	f_l&g_l\\
	-g_{-l}&-f_l
	\end{pmatrix}
	\label{eq:Gamma_space}
\end{align}
with
\begin{align}
	g_l&\equiv\im\langle a_ib_{i+l-1}\rangle\ ,\label{eq:g_fun}\\
	f_l&\equiv\im\langle a_ia_{i+l}\rangle-\im\delta_{l0}=\im\langle b_{i+l}b_i\rangle-\im\delta_{l0}\ ,\label{eq:f_fun}
\end{align}
and Majorana operators $a_i=c_i^\dagger+c_i, b_i=\im(c_i^\dagger-c_i)$
\cite{Lieb1961,Caianiello1952, Barouch1970,Barouch1971,Calabrese2012}. Here 
$\langle\cdot\rangle$ denotes the expectation value for a given state $|\psi\rangle$, i.e. 
$\langle\cdot\rangle\equiv\langle\psi|\cdot|\psi\rangle$.
Since, due to translational invariance,
\begin{align}
	g_l&=\frac{\im}{N}\sum_ke^{-\im k(l-1)}\langle b_ka_{-k}\rangle
	\equiv\frac{1}{N}\sum_ke^{-\im kl}\hat g_k\ ,\label{eq:g_k}\\
	f_l&=\frac{\im}{N}\sum_ke^{-\im kl}\langle a_ka_{-k}\rangle
	\equiv\frac{1}{N}\sum_ke^{-\im kl}\hat f_k\ ,\label{eq:f_k}
\end{align}
where $a_k=\frac{1}{\sqrt{N}}\sum_le^{-\im kl}a_l$ and $b_k=\frac{1}{\sqrt{N}}\sum_le^{-\im kl}b_l$, 
the Toeplitz matrix $\Gamma_{ij}$ is fully determined by its symbol
\begin{align}
	\hat\Gamma_k=\begin{pmatrix}
		\hat f_k&\hat g_k\\-\hat g_{-k}&-\hat f_k
	\end{pmatrix}
\end{align}
via $\Gamma_l=\sum_k e^{-\im kl}\hat\Gamma_k$.

For our purposes we consider the transverse magnetisation
\begin{align}
	\langle m_x\rangle\equiv\frac{1}{N}\sum_i\langle S_i^x\rangle=-\frac12g_1
	\label{eq:tvmagndef}
\end{align}
and the longitudinal spin-spin correlation
\begin{align}
	\rho_n^{zz}\equiv\langle S_i^zS_{i+n}^z\rangle
	=\frac14\operatorname{Pf}\left[\Gamma^n\right]\ ,
	\label{eq:szszdef}
\end{align}
where $\operatorname{Pf}[\cdot]$ denotes the Pfaffian and $\Gamma^n$ is the correlation matrix
consisting of blocks $\Gamma_{ij}$ with $|i-j|<n$ (cf. eq. \eqref{eq:Gamma_space}). 
Moreover, we will study the
entanglement entropy $S_n$ of a strip $A_n$ of $n$ adjacent spins with the rest of the system,
which is given by
\begin{align}
	S_n\equiv\operatorname{Tr}\left[\rho_{A_n}\ln\left(\rho_{A_n}\right)\right]=\sum_l\frac{1+\nu_l}{2}
	\label{eq:ent_def}
\end{align}
where $\rho_{A_n}$ is the reduced density matrix of the subsystem $A_n$ and
$\nu_l$ are the eigenvalues of $\Gamma^n$ \cite{Vidal2002,Latorre2004}.

In the following we will be interested in time evolution which is induced by quenching the magnetic
field $g$ at $t=0$. This means the initial state $|\psi_0\rangle$ is the ground state of the Hamiltonian 
$H(g_0)$ and 
for $t>0$ the time evolution is driven by a Hamiltonian $H(g)$ with $g\neq g_0$.
To compute the time evolution for this protocol it is convenient
to introduce operators
\begin{align}
	\vec\Omega_i\equiv\begin{pmatrix}\omega_i^+\\\omega_i^-\end{pmatrix}
	\equiv\sqrt2R^y(\pi/2)\begin{pmatrix}c_i^\dagger\\c_i\end{pmatrix}
\end{align}
in terms of which the correlators \eqref{eq:g_fun} and
\eqref{eq:f_fun} are
$\im\langle a_ia_j\rangle=\im\langle\omega_i^+\omega_j^+\rangle$ and 
$\im\langle a_ib_j\rangle=-\langle\omega_i^+\omega_j^-\rangle$\ .
$\Gamma^n$ is then fully determined by the correlation matrix
\begin{align}
	\langle\vec\Omega_k\vec\Omega_k^\dagger\rangle_t=
	\begin{pmatrix}
	\langle \omega_k^+\omega_{-k}^+\rangle_t&-\langle \omega_k^+\omega_{-k}^-\rangle_t\\
	\langle \omega_k^- \omega_{-k}^+\rangle_t&-\langle \omega_k^- \omega_{-k}^-\rangle_t
	\end{pmatrix}\ .
\end{align}
where $\langle\cdot\rangle_t$ is the expectation value with respect to the time evolved state
$|\psi(t)\rangle$. For the abovementioned quench the 
expectation values with $|\psi(t)\rangle=\exp\left(-\im H(g)t\right)|\psi_0\rangle$ can be 
evaluated \cite{Sengupta2004}, yielding 
\begin{align}
	\langle\vec\Omega_k\vec\Omega_k^\dagger\rangle_t=
	\frac{1}{2}\tilde U_k(t)\left(\sigma^z+1\right)\tilde U_k(t)^\dagger\ ,
\end{align}
where 
$\tilde U_k(t)=\sqrt2R^y\left(\frac{\pi}{2}\right)R^x(\theta_k^g)R^z(2\epsilon_k^gt)R^x(\phi_k^{g,g_0})$
with $R^\alpha(\phi)$ as defined in eq. \eqref{eq:unitaries} 
and $\phi_k^{g,g_0}\equiv\theta_k^g-\theta_k^{g_0}$. In the following we will employ straightforward
generalisations of this formalism for situations of imperfect effective time reversal, generally yielding
coefficients $\Sigma_\alpha^k(t)\equiv\Sigma_\alpha^k(t,g_0,g,\ldots)$ with which
\begin{align}
	\langle\vec\Omega_k\vec\Omega_k^\dagger\rangle_t
	=\idop+\sum_{\alpha\in\{x,y,z\}}\Sigma_\alpha^k(t)\sigma^\alpha\ ,
	\label{eq:corr_mat}
\end{align}
where $\sigma^\alpha$ denote the Pauli matrices.
Although derived straightforwardly, the expressions for $\Sigma_\alpha^k(t)$ become very 
lengthy for
the time reversal protocols under consideration in this work. The full expressions can be found in
the appendix.

\section{Quantifying initial state resemblance}
In the following we will study the resemblance of a time evolved state to the initial state when
different kinds of imperfect effective time reversal are employed at $t=\tau$. For this purpose we
compute different time dependent quantities $X_t$, namely observables and entanglement entropy.
For $t\to\infty$ these quantities approach a stationary value $X_\infty$. However, due to the applied
time reversal protocol the deviation $|X_t-X_\infty|$ will show a distinguished maximum at 
$t_e\approx2\tau$, which we call the echo peak. We will consider the normalised echo peak height
\begin{align}
	E_\tau^*[X]=\max_{t>\tau}\left|\frac{X_{t}-X_\infty}{X_0-X_\infty}\right|
\end{align}
as measure for the initial state resemblance.

According to eqs. \eqref{eq:g_k}, \eqref{eq:f_k}, and \eqref{eq:corr_mat} the quantities of interest
will in the thermodynamic limit ($N\to\infty$) be determined by integrals 
$\int_{-\pi}^\pi dke^{-\im nk}\Sigma_\alpha^k/2\pi$, where the
time-dependent parts of $\Sigma_\alpha^k$ oscillate more and more quickly as function of $k$ with
increasing $t$. Therefore, the stationary values $X_\infty$ are given by the corresponding
integrals over only the time-independent contributions to $\Sigma_\alpha^k$ \cite{Sengupta2004}.

In what follows we discuss three different time reversal protocols, namely time reversal by
explicit sign change of the Hamiltonian, $H(g)\to-H(g+\delta g)$,
time reversal by application of a Loschmidt pulse $U_P$, $H(g)\to U_P^\dagger H(g)U_P$,
and a generalised Hahn echo protocol, $H(g)\to H(-g)$.
All three protocols yield algebraically decaying or even ever persisting echo peak heights.

\begin{figure}[!h]
\includegraphics{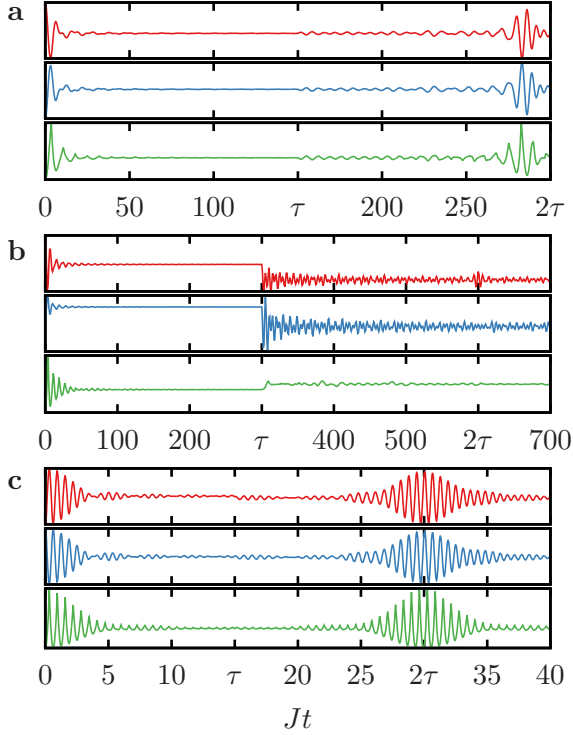}
\caption{Time evolution of the transverse magnetisation 
$\langle m_x\rangle_t$ (red curves), the longitudinal spin-spin correlation
$\langle S_i^zS_{i+1}^z\rangle_t$ (blue curves), and the 
rate function of the fidelity $l(t)=\lim_{N\to\infty}\ln(\langle\psi_0|\psi(t)\rangle)/N$ (green curves) for
the three different echo protocols: (a) by explicit sign change, 
(b) by pulse, (c) generalised Hahn echo.}
\label{fig:full_echoes}
\end{figure}

\section{Time reversal by explicit sign change}
As a first echo protocol we consider effective time reversal induced by an explicit sign change of
the Hamiltonian at time $\tau$ and a well controlled deviation in the backward evolution through 
a slight variation $\delta g$ of the magnetic field, i.e. for $t>\tau$
\begin{align}
	U(t)=\exp\left(\im H(g_\delta)(t-\tau)\right)\exp\left(-\im H(g) \tau\right)\ ,
\end{align}
where $g_\delta\equiv g+\delta g$ was introduced. The corresponding correlation matrix 
\eqref{eq:corr_mat} is given by eqs. \eqref{eq:omega_terms_x}-\eqref{eq:omega_terms_z} in the
 appendix.

In order to reliably assess how well an initial state can be recovered by imperfect effective time reversal
we choose initial states, which exhibit distinguishable expectation values of some observables. 
These are ground states of $H(g)$ for $g=0$ or $g\gg 1$, respectively, which show large spin-spin
correlations.

Under the time reversal protocol described above the energy spectrum $\epsilon_k^{g_\delta}$ 
is deformed as compared to $\epsilon_k^g$ and, consequently, the
quasiparticle velocities, $v_k=\frac{d\epsilon_k}{dk}$, during forward and backward evolution can
differ. Therefore, the closest resemblance of the time evolved state to the initial state does not
necessarily occur at $t=2\tau$. This becomes evident in the exemplary time evolution 
displayed in fig. \ref{fig:full_echoes}a.
In addition to observables fig. \ref{fig:full_echoes}a shows
the time evolution of the rate function of the fidelity, 
$l(t)=\lim_{N\to\infty}\ln(|\langle\psi_0|\psi(t)\rangle|^2)/N$. A minimum of this quantity corresponds to
a large overlap of the time evolved state with the initial state for finite $N$. Note as an aside that the 
time evolution exhibits dynamical quantum phase transitions, which aroused a lot of interest recently
\cite{Heyl2013}, 
in the forward as well as in the backward evolution.

\begin{figure}[!h]
\includegraphics{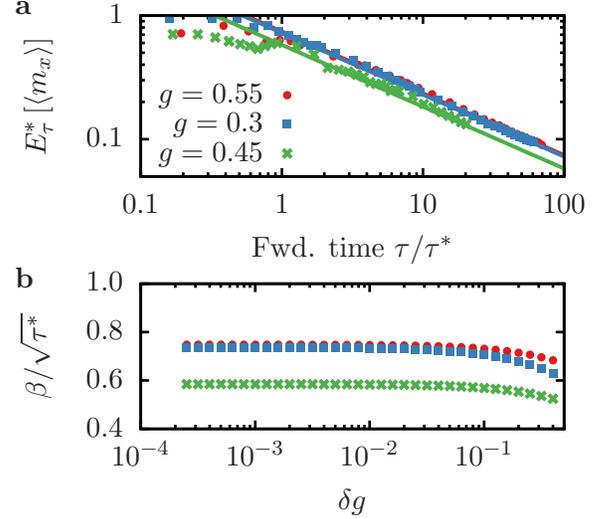}
\caption{(a) Echo peak height of the transverse magnetisation for three different quenches.
The dots are exact results, the lines are the asymptotes $\propto\tau^{-1/2}$ given by 
eq. \eqref{eq:algdecay}.
(b) Estimation of the echo peak height at the onset of the algebraic decay 
based on the stationary phase approximation. 
The echo protocol parameters are $g_0=1$ and $\delta g=0.02$.}
\label{fig:tmagnspa}
\end{figure}

Let us first consider echoes in the transverse magnetisation. 
For this observable a stationary phase approximation reveals an
algebraic decay of the echo peak height with
\begin{align}
	|\langle m_x\rangle_{t_e}-\langle m_x\rangle_\infty|
	\approx\kappa_{k^*}(\tau)\beta_{k^*}^{g,g_0,g_\delta}\tau^{-1/2}\ ,\label{eq:algdecay}
\end{align}
where $\kappa_{k^*}(\tau)=\cos(2(\epsilon_{k^*}^g- \nu_{k^*}^{g,g_\delta}\epsilon_{k^*}^{g_\delta})\tau+\pi/4)$,
\begin{align}
	\beta_{k^*}^{g,g_0,g_\delta}=&
	\frac{\zeta_{k^*}^{g,g_\delta,g_0}}{2\sqrt{\pi}\left|\xi_{k^*}^{g,g_\delta}\right|^{1/2}}\ ,\\
	\zeta_{k}^{g,g_\delta,g_0}\equiv&\sin\theta_{k^*}^{g_\delta}\sin\phi_{k^*}^{g,g_0}\frac{\cos\phi_{k^*}^{g_\delta,g}+1}{2}\ ,\\
	\xi_{k^*}^{g,g_\delta}\equiv&\left.\frac{d^2}{dk^2}(\epsilon_k^g- \nu_{k^*}^{g,g_\delta}\epsilon_k^{g_\delta})\right|_{k=k^*}\ ,
\end{align}
and the echo peak time is $t_e=(1+\nu_{k^*}^{g,g_\delta})\tau$ with
\begin{align}
	k^*=\argmax{k^*}\left|\beta_{k^*}^{g,g_0,g_\delta}\right|\ .
\end{align}
From the stationary phase approximation the onset of the algebraic decay can be expected at 
$\tau\approx\tau^*$ with 
$\tau^*=\left.\frac{d}{dk}\zeta_k^{g,g_\delta,g_0}\right|_{k=k^*}/\xi_{k^*}^{g,g_\delta}$. On this
timescale the oscillating factor varies so slowly that it can be approximated with 
$\kappa_{k^*}(\tau)\approx1$.
A detailed derivation of this result is given in the appendix. 
Fig. \ref{fig:tmagnspa}a shows the decay of the echo peak height of the transverse magnetisation as
a function of the forward time $\tau$ for three different quenches starting in the paramagnetic phase.
In all cases the initial magnetisation is almost 
perfectly recovered for forward times $\tau\lesssim\tau^*$, whereas it decays algebraically for
$\tau\gg\tau^*$. Moreover, the evaluation of $\beta_{k^*}^{g,g_0,g_\delta}$ and $\tau^*$ as function of
the deviation $\delta g$ yields $\beta_{k^*}^{g,g_0,g_\delta}\propto\delta g^{-1/2}$ and $\tau^*\propto
\delta g^{-1}$ for a wide range of perturbation strengths $\delta g$. As a result, 
$\beta_{k^*}^{g,g_0,g_\delta}(\tau^*)^{-1/2}$ is almost constant (cf. fig. \ref{fig:tmagnspa}b)
meaning that generally a very 
pronounced echo peak can be expected until the onset of the algebraic decay and the height of which
is independent of the imperfection in the backwards evolution. Hence, the echo peak decay is ultimately
induced by dephasing due to the deformed spectrum in the backwards evolution.

\begin{figure}[!h]
\includegraphics{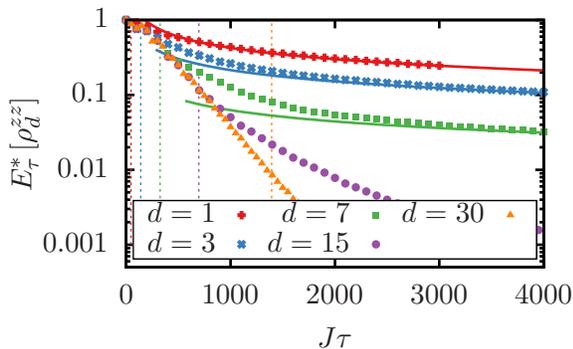}
\caption{Echo peak height of the longitudinal spin-spin correlation for
different distances $d$ with quench parameters $g_0=0, g=1,\delta g=0.05$.
The dots are exact results, the solid lines are $\propto\tau^{-1/2}$ and the dashed lines mark the
forward times $\tau=d/\tilde v_\text{max}$ for the different $d$, respectively. 
}
\label{fig:szsz_ent}
\end{figure}

Fig. \ref{fig:szsz_ent} shows the longitudinal spin-spin correlation $\rho_d^{zz}$ computed 
according to eq. \eqref{eq:szszdef} for different distances $d$. 
For this quantity we also observe an algebraic decay of the echo peak height, 
$E_\tau^*[\rho_d^{zz}]\propto\tau^{-1/2}$ for large $\tau$. 
However, before the onset of the algebraic
decay there is a distance-dependent regime of exponential-looking decay, which increases with 
increasing spin-separation $d$. Similar behaviour is known for the decay of correlation 
functions after a simple quench without time reversal \cite{Calabrese2011,Calabrese2012}. 
In that case the decay law can be rigorously derived by
identifying an space-time scaling regime where $v_\text{max}t\sim d$ with $v_\text{max}$ the
maximal propagation velocity. Due to the similar algebraic structure in the echo dynamics 
we expect a similar explanation
for the intermediate regime in the decay of the echo peak of $\rho_d^{zz}$ with a different
effective velocity 
$\tilde v_\text{max}=\underset{k\in[0,\pi]}{\text{max}}\frac{d}{dk}(\epsilon_k^g- \nu_{k^*}^{g,g_\delta}\epsilon_k^{g_\delta})$. 
At late times all entries of the 
correlation matrix \eqref{eq:Gamma_space} will just like the transverse magnetisation decay 
algebraically with
exponent $-1/2$, and therefore the leading term of the Pfaffian will decay with the same power law,
which explains that $E_\tau^*[\rho_d^{zz}]\propto\tau^{-1/2}$ for $\tilde v_\text{max}\tau\gg d$.
Considering the order parameter $\langle m_z\rangle=\lim_{d\to\infty}\rho_d^{zz}$, our result implies
an exponential decay of the echo peak height for all forward times $\tau\gg1$ when starting from
an ordered state.

\begin{figure}[!h]
\includegraphics{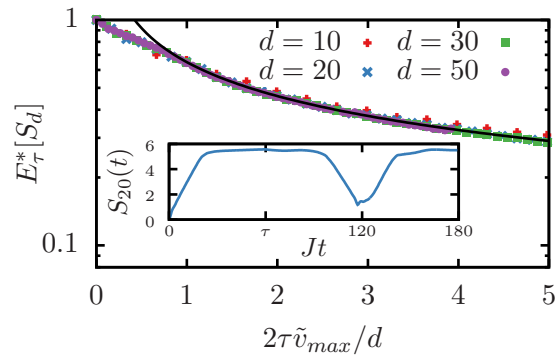}
\caption{Decay of the echo peak height of the entanglement entropy for different subsystem sizes $d$, 
$g_0=1, g=0.5,\delta g=0.05$.
After long waiting times the echo peak height decays algebraically $\propto\tau^{-1/2}$ (black line).
The inset shows an exemplary time evolution of the entanglement entropy under effective time reversal
for $d=20$ and $g_0=2.5, g=0.55,\delta g=0.05,\tau=30$.}
\label{fig:ent}
\end{figure}

Another interesting question is how far the entanglement produced during the time evolution can be
reduced again by the effective time reversal protocol.
After quenching the magnetic field the entanglement entropy increases until it saturates at
a level that is determined by the subsystem size \cite{Calabrese2005}. Therefore, an echo peak
$E_\tau^*[S_d]=1$ means that the state $|\psi(t_e)\rangle$ has the same low entanglement as the
initial state, whereas for $E_\tau^*[S_d]<1$ some additional entanglement remains.
Fig. \ref{fig:ent} displays the echo peak height of the entanglement entropy $S_n$ of 
a subsystem $A_d$ consisting of $d$ adjacent spins with the rest of the chain as defined in 
eq. \eqref{eq:ent_def}. Again we observe an initial $d$-dependent regime of 
exponential-looking decay crossing over to algebraic decay with 
$E_\tau^*[S_d]\propto\tau^{-1/2}$ for $\tau>d/2\tilde v_\text{max}$.

\section{Time reversal by a Loschmidt pulse}
Another possibility to invert the course of the dynamics in the TFIM is the application of a pulse similar
to the $\pi$-pulses applied to the system in a Hahn echo experiment. Consider the Hamiltonian
\begin{align}
	H_P=-\alpha\sum_j\left(S_j^xS_{j+1}^y+h.c.\right)\ .\label{eq:pulseham}
\end{align}
In terms of the Jordan-Wigner fermions in momentum space this reads
\begin{align}
	H_P=2\alpha\sum_k\sin k\left(c_k^\dagger c_{-k}^\dagger+c_{-k} c_k\right)
\end{align}
and the structure does not change under Bogoliubov rotation, since
$c_k^\dagger c_{-k}^\dagger+c_{-k} c_k
=\lambda_k^\dagger \lambda_{-k}^\dagger+\lambda_{-k} \lambda_k$.
The time evolution operator for this Hamiltonian in the diagonal basis is
\begin{align}
	e^{-\im H_Pt}&=\prod_k\left[\cos(2\alpha t\sin k)\right.\nonumber\\
	&\quad\quad\quad\left.
	-\im\sin(2\alpha t\sin k)\left(\lambda_k^\dagger \lambda_{-k}^\dagger
	+\lambda_{-k} \lambda_k\right)\right]
\end{align}
For a given pulse time $t_P$ this operator perfectly inverts the population of modes $k_n^*$ with
$2\alpha t_P\sin k_n^*=(2n+1)\pi/2, n\in\mathbb Z$, whereas the population of modes $k_n^\times$
with $2\alpha t_P\sin k_n^\times=n\pi$ remains unchanged. The population of other modes is
partially inverted. Thereby, the evolution of the system under the Hamiltonian \eqref{eq:pulseham} for
a pulse time $t_p$ leads to an imperfect effective time reversal, i.e.
$e^{\im H_Pt_P} e^{-\im H \tau}e^{-\im H_Pt_P}=e^{\im (H+\epsilon V)\tau}$,
by an imperfect inversion of the mode occupation.

Note that in the notation introduced above the pulse operator in one $k$-sector is
$\tilde U_k^P(2\alpha t)\equiv R^x(4\alpha t\sin k)^\dagger$.
The result for the corresponding correlation matrix  to this echo protocol is given in eqs. 
\eqref{eq:sigma_x_pulse}-\eqref{eq:sigma_z_pulse} in the
appendix.

\begin{figure}[!h]
\includegraphics{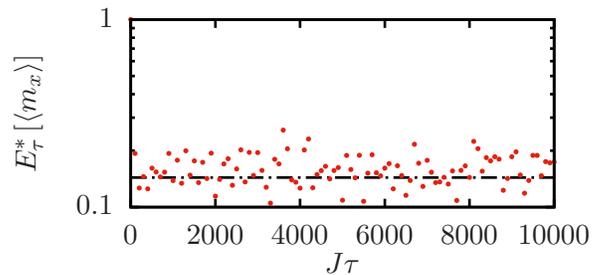}
\caption{Echo peak heights for the transverse magnetisation $m_x$ when applying the pulse 
Hamiltonian
\eqref{eq:pulseham} for effective time reversal. Due to absence of dephasing the echo peak height 
(red dots) does not decay but
oscillates around the value given by eq. \eqref{eq:res_peak_height} (dashed line).
The parameters of the corresponding pulse echo protocol are $g_0=1$,$g=0.3$, and $\alpha t_P=25$.}
\label{fig:pulse_echo}
\end{figure}

With this protocol the forward and backward time evolution are generated by the same Hamiltonian;
hence, the echo peak after time reversal at time $\tau$ appears at $t_e=2\tau$, which can be seen
in the exemplary time evolution in fig. \ref{fig:full_echoes}b. In particular, we 
find that at $t_e=2\tau$ the transverse magnetisation can be split into three parts,
\begin{align}
	\langle m_x\rangle_{t_e=2\tau}=\langle m_x\rangle_\infty+\langle m_x\rangle_E
	+\langle m_x\rangle_{\tau}\ ,\label{eq:pulse_mx}
\end{align}
where $\langle m_x\rangle_\infty$ is the stationary value reached at $t\to\infty$, 
$\langle m_x\rangle_E$ is an additional $\tau$-independent contribution, 
and $\langle m_x\rangle_{\tau}$ are the time-dependent contributions, which vanish for $\tau\to\infty$.
This means we find an echo peak at $t_e=2\tau$, which never decays. The residual peak height is
given by
\begin{align}
	\langle m_x\rangle_E=\frac12\int_{-\pi}^\pi \frac{dk}{2\pi}\sin\phi_k^{g,g_0}
	\Big(1-\cos\big(4\alpha t_P\sin k\big)\Big)
	\label{eq:res_peak_height}
\end{align}
An example of this ever
persisting echo is depicted in fig. \ref{fig:pulse_echo}, where the dots show the exact result and the
dashed line shows the echo peak height expected when only considering the time-independent
contributions in eq. \eqref{eq:pulse_mx}.

\section{Time reversal by generalised Hahn echo}
The Hamiltonian of the TFIM \eqref{eq:tfim_ham} allows for an echo protocol very similar to the way the
effective time reversal is induced in a Hahn echo experiment \cite{Hahn1980}, namely by a sign 
inversion of the
Zeeman term. For large magnetic fields changing the sign of the field can be considered an
effective time reversal with small imperfection given by the Ising term, 
$\epsilon V=2J\sum_iS_i^zS_{i+1}^z$. In contrast to the original
Hahn echo setup, where an initial magnetisation decays due to field inhomogeneities, the decay in 
the TFIM will be due to the coupling of the physical degrees of freedom. 

The stationary phase analysis for the echo dynamics under this protocol unveils that it combines two
properties of the previously discussed protocols. Since the switching of the magnetic field 
corresponds
to a shift of the energy spectrum, $\epsilon_k^g=\epsilon_{k+\pi}^{-g}$, the quasiparticle velocities are
perfectly inverted, yielding echo peaks at $t_e=2\tau$; nevertheless, for large $\tau$ 
the echo peak height decays algebraically with exponent $-1/2$. A detailed derivation is given
in the appendix.


\section{Relation to thermalisation}
The fact that it is well possible to produce pronounced echoes in the TFIM also after long waiting
times matches the absence of thermalisation in the conventional sense. The dynamics of the
system is constrained by infinitely many integrals of motion, which keep a lot of information about the
initial state. Especially, these integrals of motion determine the stationary value of local observables
through the corresponding generalised Gibbs ensemble (GGE) \cite{Calabrese2011}, i.e. the reduced
density matrix of a 
strip of length $l$, $\rho_l(t)$, converges to a density matrix given by a GGE, $\rho_{GGE,l}$, 
for $t\to\infty$. Note that  the distance of both density matrices decreases as 
$\mathcal D(\rho_l(t),\rho_{GGE,l})\propto l^2t^{-3/2}$\cite{Fagotti2013}, 
whereas the expectation value of
an observable at $t=2\tau$ is determined by
$\langle O\rangle_{2\tau}=\text{tr}[O(\tau)\rho(\tau)]$. In the latter expression 
$\rho(\tau)=e^{-iH\tau}|\psi_0\rangle\langle\psi_0|e^{iH\tau}$ approaches a GGE
as mentioned above but $O(\tau)=e^{-i(H+\epsilon V)\tau}Oe^{i(H+\epsilon V)\tau}$ 
becomes increasingly non-local.
Therefore, the fact that echoes are possible after arbitrarily long waiting times does not contradict the
convergence to a GGE.

\section{Discussion}
We proposed a definition of irreversibility based on the decay of observable echoes under
imperfect effective time reversal and presented different ways to induce the time reversal 
in the TFIM. As a result we
find an algebraic decay of the echo peak height after long forward times for all observables
under consideration due to dephasing whenever the
imperfection comes along with a deformation of the energy spectrum. 
In the case of an unchanged
spectrum during forward and backward evolution there is a residual contribution to the echo peak,
which never decays.

Based on these results we conclude that the dynamics in the TFIM can be considered well
reversible. This finding matches the fact that the TFIM has an infinite number of integrals of motion,
which preserve a lot of information about the initial state throughout the
course of the dynamics and also prevent the equilibration to a conventional Gibbs ensemble.

An important point of future work will be to understand the dynamics of non-quadratic Hamiltonians 
under imperfect effective time reversal. 
Work along these lines is in progress.






\begin{acknowledgements}
The authors thank S. Sondhi for the interesting suggestion to study the entanglement entropy
under effective time reversal as well as M. Medvedyeva, M. Heyl, M. Fagotti, 
and D. Fioretto for helpful discussions and N. Abeling for proofreading the manuscript.
This work was supported through SFB
1073 (project B03) of the Deutsche Forschungsgemeinschaft
(DFG) and the Studienstiftung des Deutschen Volkes.
For the numerical computations the Armadillo library \cite{armadillo} was used.
\end{acknowledgements}
\bibliography{refs}




\clearpage
\newpage
\onecolumngrid
\appendix
\section{Computing echo time evolution in the TFIM}
As mentioned in the main text the time evolution of all observables of interest is essentially determined
by the correlation matrix
\begin{align}
	\langle\vec\Omega_k\vec\Omega_k^\dagger\rangle_t=
	\begin{pmatrix}
	\langle \omega_k^+\omega_{-k}^+\rangle_t&-\langle \omega_k^+\omega_{-k}^-\rangle_t\\
	\langle \omega_k^- \omega_{-k}^+\rangle_t&-\langle \omega_k^- \omega_{-k}^-\rangle_t
	\end{pmatrix}\ ,
\end{align}
which is for the case of quenching from the ground state determined by
\begin{align}
	\langle\vec\Omega_k\vec\Omega_k^\dagger\rangle_t=
	\frac{1}{2}\tilde U_k(t)\left(\sigma^z+1\right)\tilde U_k(t)^\dagger
\end{align}
with 
$\tilde U_k(t)=\sqrt2R^y\left(\frac{\pi}{2}\right)R^x(\theta_k^g)R^z(2\epsilon_k^gt)R^x(\phi_k^{g,g_0})$ 
the time evolution operator of the corresponding $k$-sector in expressed in the basis
of the initial Hamiltonian \cite{Sengupta2004}.
Recall the definition
\begin{align}
	R^\alpha(\phi)=\idop\cos\frac{\phi}{2}+\im\sigma^\alpha\sin\frac{\phi}{2}\ ,\quad \alpha\in\{x,y,z\}\ .
\end{align}
The correlation matrices can generally be written
as
\begin{align}
	\langle\vec\Omega_k\vec\Omega_k^\dagger\rangle_t
	=\idop+\sum_{\alpha=1}^3\Sigma_\alpha^k(t)\sigma^\alpha
	\label{eq:sigma_def}
\end{align}
with suited coefficients $\Sigma_\alpha^k(t)$, $\alpha\in\{x,y,z\}$, and the Pauli matrices 
$\sigma^\alpha$.
The formalism for a simple quench is straightforwardly
generalised for the different echo protocols discussed in the main text. In the following subsections we
give the corresponding coefficient functions $\Sigma_\alpha^k$.

\subsection{$\Sigma_\alpha$ for echoes through explicit sign change}
For the time reversal by explicit sign change, where the Hamiltonian is
switched from $H(g)$ to $-H(g+\delta g)$ at $t=\tau$, $\tilde U_k(t)$ generalises to
\begin{align}
\tilde U_k(t)=&\sqrt2R^y\left(\frac{\pi}{2}\right)\left\{
	\begin{matrix}
	R^x(\theta_k^g)&,\quad t<\tau\\
	R^x(\theta_k^{g_\delta})R^z(-2\epsilon_k^{g_\delta}(t-\tau))^\dagger R^x(\phi_k^{g_\delta,g})^\dagger&,\quad t>\tau
	\end{matrix}\right\} R^z(2\epsilon_k^gt)^\dagger R^x(\phi_k^{g,g_0})^\dagger
\end{align}
yielding
\begin{align}
	\Sigma_x^k=&-\left[\left(\cos\phi_k^{g,g_0}\sin\phi_k^{g_\delta,g}+\sin\phi_k^{g,g_0}\cos\phi_k^{g_\delta,g}\cos(2\epsilon_k^g\tau)\right)\cos(2\epsilon_k^{g_\delta}(t-\tau))\right.\nonumber\\
		&\hspace{4cm}\left.+\sin\phi_k^{g,g_0}\sin(2\epsilon_k^g\tau)\sin(2\epsilon_k^{g_\delta}(t-\tau))\right]\sin\theta_k^{g_\delta}\nonumber\\
		&-\left[\cos\phi_k^{g,g_0}\cos\phi_k^{g_\delta,g}-\sin\phi_k^{g,g_0}\sin\phi_k^{g_\delta,g}\cos(2\epsilon_k^g\tau)\right]\cos\theta_k^{g_\delta},\label{eq:omega_terms_x}\\
	\Sigma_y^k=&-\left[\left(\cos\phi_k^{g,g_0}\sin\phi_k^{g_\delta,g}+\sin\phi_k^{g,g_0}\cos\phi_k^{g_\delta,g}\cos(2\epsilon_k^g\tau)\right)\cos(2\epsilon_k^{g_\delta}(t-\tau))\right.\nonumber\\
	&\hspace{4cm}\left.+\sin\phi_k^{g,g_0}\sin(2\epsilon_k^g\tau)\sin(2\epsilon_k^{g_\delta}(t-\tau))\right]\cos\theta_k^{g_\delta}\nonumber\\
		&+\left[\cos\phi_k^{g,g_0}\cos\phi_k^{g_\delta,g}-\sin\phi_k^{g,g_0}\sin\phi_k^{g_\delta,g}\cos(2\epsilon_k^g\tau)\right]\sin\theta_k^{g_\delta}\ ,\label{eq:omega_terms_y}\\
	\Sigma_z^k=&\sin\phi_k^{g,g_0}\sin(2\epsilon_k^g\tau)\cos(2\epsilon_k^{g_\delta}(t-\tau))\nonumber\\
		&-\left(\cos\phi_k^{g,g_0}\sin\phi_k^{g_\delta,g}+\sin\phi_k^{g,g_0}\cos\phi_k^{g_\delta,g}\cos(2\epsilon_k^g\tau)\right)\sin(2\epsilon_k^{g_\delta}(t-\tau))\ .\label{eq:omega_terms_z}
\end{align}
for $t>\tau$. For $t\to\infty$ and $\tau<t$ the stationary value after application of the time reversal 
is determined by the time-independent contributions
\begin{align}
	\Sigma_x^k=&-\cos\phi_k^{g,g_0}\cos\phi_k^{g,g_\delta}\cos\theta_k^g\ ,
	\label{eq:omega_terms_fwd_lt_x}\\
	\Sigma_y^k=&\cos\phi_k^{g,g_0}\cos\phi_k^{g,g_\delta}\sin\theta_k^g\ ,
	\label{eq:omega_terms_fwd_lt_y}\\
	\Sigma_z^k=&0\ .\label{eq:omega_terms_fwd_lt_z}
\end{align}

\subsection{$\Sigma_\alpha$ for echoes through pulse Hamiltonian} 
In the case of time reversal by application of a pulse operator 
$\tilde U_k^P(2\alpha t)= R^x(4\alpha t\sin k)^\dagger$ we obtain
\begin{align}
	\Omega(t)=&\sqrt2R^y\left(\frac{\pi}{2}\right)R^x(\theta_k^g)
	\left\{
	\begin{matrix}
	R^z(2\epsilon_k^gt)^\dagger&,\quad t<\tau\\
	R^x(4\alpha t_p\sin k) R^z(2\epsilon_k^g(t-\tau))^\dagger R^x(4\alpha t_p\sin k)^\dagger R^z(2\epsilon_k^g\tau)^\dagger&,\quad t>\tau
	\end{matrix}\right\}\nonumber\\
	&\cdot R^x(\phi_k^{g,g_0})^\dagger\begin{pmatrix}\gamma_k\\\gamma_{-k}^\dagger\end{pmatrix}
\end{align}
In terms of eq. \eqref{eq:sigma_def} we get for $t>\tau$
\begin{align}
	\Sigma_x^k=&\left(A\cos\chi_k^{\alpha t_p}+B\sin\chi_k^{\alpha t_p}\right)\sin\theta_k^g
		-\left(-A\sin\chi_k^{\alpha t_p}+B\cos\chi_k^{\alpha t_p}\right)\cos\theta_k^g\ ,
		\label{eq:sigma_x_pulse}\\
	\Sigma_y^k=&\left(A\cos\chi_k^{\alpha t_p}+B\sin\chi_k^{\alpha t_p}\right)\cos\theta_k^g
		+\left(-A\sin\chi_k^{\alpha t_p}+B\cos\chi_k^{\alpha t_p}\right)\sin\theta_k^g\ ,
		\label{eq:sigma_y_pulse}\\
	\Sigma_z^k=&\sin\phi_k^{g,g_0}\sin(2\epsilon_k(g)\tau)\cos(2\epsilon_k(g)(t-\tau))\nonumber\\
			&+\left(\sin\phi_k^{g,g_0}\cos(2\epsilon_k(g)\tau)\cos(\chi_k^{\alpha t_p})-\cos\phi_k^{g,g_0}\sin(\chi_k^{\alpha t_p})\right)\sin(2\epsilon_k(g)(t-\tau))\ ,
			\label{eq:sigma_z_pulse}
\end{align}
where $\chi_k^{\alpha t_p}=4\alpha t_p\sin k$ was introduced and
\begin{align}
	A=&\sin\phi_k^{g,g_0}\sin(2\epsilon_k(g)\tau)\sin(2\epsilon_k(g)(t-\tau))\nonumber\\
	&-\left(\sin\phi_k^{g,g_0}\cos(2\epsilon_k(g)\tau)\cos\chi_k^{\alpha t_p}-\cos\phi_k^{g,g_0}\sin\chi_k^{\alpha t_p}\right)\cos(2\epsilon_k(g)(t-\tau))\ ,\\
	B=&\sin\phi_k^{g,g_0}\cos(2\epsilon_k(g)\tau)\sin\chi_k^{\alpha t_p}+\cos\phi_k^{g,g_0}\cos\chi_k^{\alpha t_p}
\end{align}

\subsection{$\Sigma_\alpha$ for generalised Hahn echo}
In the generalised Hahn echo protocol the time reversal is at $t=\tau$ induced by a sign change of the 
magnetic field, $g\to-g$. In this case the time evolution operator is
\begin{align}
\tilde U_k(t)=&\sqrt2R^y\left(\frac{\pi}{2}\right)\left\{
	\begin{matrix}
	R^x(\theta_k^g)&,\quad t<\tau\\
	R^x(\theta_k^{-g})R^z(2\epsilon_k^{-g}(t-\tau))^\dagger R^x(\phi_k{-g,g})^\dagger&,\quad t>\tau
	\end{matrix}\right\}R^z(2\epsilon_k^gt)^\dagger R^x(\phi_k^{g,g_0})^\dagger
\end{align}
and the corresponding coefficients for the correlation matrix are
\begin{align}
	\Sigma_x^k=&-\left[\left(\cos\phi_k^{g,g_0}\sin\phi_k^{-g,g}+\sin\phi_k^{g,g_0}\cos\phi_k^{-g,g}\cos(2\epsilon_k^g\tau)\right)\cos(2\epsilon_k^{-g}(t-\tau))\right.\nonumber\\
		&\hspace{4cm}\left.-\sin\phi_k^{g,g_0}\sin(2\epsilon_k^g\tau)\sin(2\epsilon_k^{-g}(t-\tau))\right]\sin\theta_k^{-g}\nonumber\\
		&-\left[\cos\phi_k^{g,g_0}\cos\phi_k^{-g,g}-\sin\phi_k^{g,g_0}\sin\phi_k^{-g,g}\cos(2\epsilon_k^g\tau)\right]\cos\theta_k^{-g},\label{eq:omega_terms_x_hahn}\\
	\Sigma_y^k=&-\left[\left(\cos\phi_k^{g,g_0}\sin\phi_k^{-g,g}+\sin\phi_k^{g,g_0}\cos\phi_k^{-g,g}\cos(2\epsilon_k^g\tau)\right)\cos(2\epsilon_k^{-g}(t-\tau))\right.\nonumber\\
	&\hspace{4cm}\left.-\sin\phi_k^{g,g_0}\sin(2\epsilon_k^g\tau)\sin(2\epsilon_k^{-g}(t-\tau))\right]\cos\theta_k^{-g}\nonumber\\
		&+\left[\cos\phi_k^{g,g_0}\cos\phi_k^{-g,g}-\sin\phi_k^{g,g_0}\sin\phi_k^{-g,g}\cos(2\epsilon_k^g\tau)\right]\sin\theta_k^{-g}\ ,\label{eq:omega_terms_y_hahn}\\
	\Sigma_z^k=&\sin\phi_k^{g,g_0}\sin(2\epsilon_k^g\tau)\cos(2\epsilon_k^{-g}(t-\tau))\nonumber\\
		&+\left(\cos\phi_k^{g,g_0}\sin\phi_k^{-g,g}+\sin\phi_k^{g,g_0}\cos\phi_k^{-g,g}\cos(2\epsilon_k^g\tau)\right)\sin(2\epsilon_k^{-g}(t-\tau))\ .\label{eq:omega_terms_z_hahn}
\end{align}

\section{Stationary phase approximation for the decay after long waiting times}
\subsection{Transverse magnetisation for time reversal by explicit sign change}
The echo in the transverse magnetisation is given by
\begin{align}
	\langle m_x\rangle_{t_e}=-\int_{-\pi}^\pi\frac{dk}{4\pi}\left(\Sigma_x^k(t_e)+\im\Sigma_y^k(t_e)\right)
\end{align}
(cf. eq. \eqref{eq:tvmagndef} in the main text). Since $\theta_k^g$ is an odd function of $k$ and so is $\phi_k^{g,g'}$, whereas $\epsilon_k^g$ is even,the imaginary part does not contribute,
$\int_{-\pi}^\pi dk\Sigma_y^k=0$. Thus,
\begin{align}
	\langle m_x\rangle_{t_e}=&-\int_{-\pi}^\pi\frac{dk}{4\pi}\Sigma_x^k\\
	=&\langle m_x\rangle_\infty
	+\int\frac{dk}{4\pi}\sin\theta_k^{g_\delta}\cos\phi_k^{g,g_0}\sin\phi_k^{g_\delta,g}\cos(2\epsilon_k^{g_\delta}\nu\tau)
	+\int\frac{dk}{4\pi}\sin\theta_k^{g_\delta}\sin\phi_k^{g,g_0}\cos\phi_k^{g_\delta,g}\cos(2\epsilon_k^g\tau)\cos(2\epsilon_k^{g_\delta}\nu\tau)\nonumber\\
	&\hspace{1cm}+\int\frac{dk}{4\pi}\sin\phi_k^{g,g_0}\sin(2\epsilon_k^g\tau)\sin(2\epsilon_k^{g_\delta}\nu\tau)\sin\theta_k^{g_\delta}
	-\int\frac{dk}{4\pi}\cos\theta_k^{g_\delta}\sin\phi_k^{g,g_0}\sin\phi_k^{g_\delta,g}\cos(2\epsilon_k^g\tau)\ ,
\end{align}
where $\nu\equiv(t_e-\tau)/\tau$ was introduced.
This expression contains two types of integrands, namely two integrals including only a single 
trigonometric function of time and two integrals including a product of two trigonometric functions of time
with slightly differing spectra $\epsilon_k^{g/g_\delta}$. 

Consider the first type. For large $\tau$ the integrands become highly oscillatory and 
the main contribution to the integral is given by the stationary points $k^*$ of $\epsilon_k^g$,
\begin{align}
	\left.\frac{d\epsilon_k^g}{dk}\right|_{k=k^*}=0\Rightarrow k^*=0,\pm\pi\ .
\end{align}
At the stationary points $\phi_{k^*}=\theta_{k^*}=0$ and expanding the respective
test functions $f(k)$ around the stationary points yields
\begin{align}
	\int_{-\pi}^\pi\frac{dk}{2\pi}f(k)\cos(2\epsilon_k\tau)\approx&\sum_{k^*\in\{0,\pi\}}\int_{-\infty}^\infty\frac{dk}{2\pi}\frac{f''(k^*)}{2}(k-k^*)^2\cos(\epsilon_{k^*}''(k-k^*)^2\tau)\\
	=&\sum_{k^*\in\{0,\pi\}}\frac{1}{\tau^{3/2}}\int_{-\infty}^\infty\frac{dq}{2\pi}\frac{f''(k^*)}{2}q^2\cos(\epsilon_{k^*}''q^2)\propto\frac{\cos(2\epsilon_{k^*}\tau+\pi/4)}{\tau^{3/2}}
\end{align}
Thus, at long forward times these parts give contributions with a decay law $\propto\tau^{-3/2}$.

Now consider the terms including products of trigonometric functions of time. 
Here we can get rid of the products via the identities
\begin{align}
	2\cos(2\epsilon_k^g\tau)\cos(2\epsilon_k^{g_\delta}\nu\tau)
	=&\cos(2(\epsilon_k^g+\nu\epsilon_k^{g_\delta})\tau)+\cos(2(\epsilon_k^g-\nu\epsilon_k^{g_\delta})\tau)\ ,\\
	2\sin(2\epsilon_k^g\tau)\sin(2\epsilon_k^{g_\delta}\nu\tau)
	=&-\cos(2(\epsilon_k^g+\nu\epsilon_k^{g_\delta})\tau)+\cos(2(\epsilon_k^g-\nu\epsilon_k^{g_\delta})\tau)\ .
\end{align}
Allowing for $0<\nu\neq1$ relevant saddle points for the above integrals are determined by
\begin{align}
	\frac{d}{dk}\left.\left(\epsilon_k^g\pm \nu\epsilon_k^{g_\delta}\right)\right|_{k=k^*}=\left.\left(\left(\frac{g}{\epsilon_k^g}\pm \nu\frac{g_\delta}{\epsilon_k^{g_\delta}}\right)\sin k\right)\right|_{k=k^*}=0
\end{align}
yielding $k^*=0,\pi$ for both signs, and additionally
\begin{align}
	k_-^*=\pm\arccos\left(\frac{\nu^2g_\delta^2(1+g^2)-g^2(1+g_\delta^2)}{2g\nu^2g_\delta^2-2g_\delta g^2}\right)
\end{align}
for the negative sign. These additional saddle points coincide with $k^*=0,\pi$ for
\begin{align}
	\nu_0=\frac{g|1+g_\delta|}{g_{\delta}|1+g|}\ ,\quad \nu_\pi=\frac{g|1-g_\delta|}{g_{\delta}|1-g|}\ ,
\end{align}
respectively. In the search of an echo peak $\nu\in[\nu_0,\nu_\pi]$ can be tuned to create any saddle point $k^*\in[0,\pi]$ via
\begin{align}
	\nu_{k^*}^{g,g_\delta}=\frac{g\epsilon_{k^*}^{g_\delta}}{g_\delta\epsilon_{k^*}^g}\ .
\end{align}
Thus, for sufficiently long forward times with $t_e=(1+\nu)\tau$ and considering the equal contributions
of both stationary points
\begin{align}
	\langle m_x\rangle_{t_e}-\langle m_x\rangle_\infty\approx&
	-\int_{-\pi}^\pi\frac{dk}{2\pi}
	\left[\sin\theta_k^{g_\delta}\sin\phi_k^{g,g_0}\frac{\cos\phi_k^{g_\delta,g}+1}{2}\right]
	\cos\left(2(\epsilon_k^g- \nu\epsilon_k^{g_\delta})\tau\right)+\gamma\tau^{-3/2}\\
	\approx&-\beta_\nu^{g,g_0,g_\delta}\frac{\kappa_{k^*}(\tau)}{\tau^{1/2}}+\gamma\tau^{-3/2}
\end{align}
where $\kappa_{k^*}(\tau)=\cos(2(\epsilon_{k^*}^g- \nu\epsilon_{k^*}^{g_\delta})\tau+\pi/4)$ varies very slowly,
\begin{align}
	\beta_{k^*}^{g,g_0,g_\delta}=&
	\frac{\zeta_{k^*}^{g,g_\delta,g_0}}{2\sqrt{\pi}\left|\xi_{k^*}^{g,g_\delta}\right|^{1/2}}\ ,\\
	\zeta_{k}^{g,g_\delta,g_0}\equiv&\sin\theta_{k^*}^{g_\delta}\sin\phi_{k^*}^{g,g_0}\frac{\cos\phi_{k^*}^{g_\delta,g}+1}{2}\ ,\\
	\xi_{k^*}^{g,g_\delta}\equiv&\left.\frac{d^2}{dk^2}(\epsilon_k^g- \nu_{k^*}^{g,g_\delta}\epsilon_k^{g_\delta})\right|_{k=k^*}\ ,
\end{align}
and the echo peak appears at $t_e=(1+\nu_{k^*}^{g,g_\delta})\tau$, where
\begin{align}
	k^*=\argmax{k^*}\left|\beta_{k^*}^{g,g_0,g_\delta}\right|\ .
\end{align}
The stationary phase approximation is valid if the test function 
$\zeta_{k}^{g,g_\delta,g_0}\approx\zeta_{k^*}^{g,g_\delta,g_0}
+\left.\frac{d}{dk}\zeta_{k}^{g,g_\delta,g_0}\right|_{k=k^*}(k-k^*)$ 
does not vary too much on the interval $[k-\Delta k,k+\Delta k]$, where $\Delta k$ is given by the
width of the saddle point, $(\tau\xi_{k^*}^{g,g_\delta})^{-1}$, i.e. we expect the approximation to 
be good for 
$\tau>\tau^*=\left.\frac{d}{dk}\zeta_{k}^{g,g_\delta,g_0}\right|_{k=k^*}/\xi_{k^*}^{g,g_\delta}$.
Moreover, the period of $\kappa_{k^*}(\tau)$ is determined by the difference of the spectra and, hence, 
very large compared to $\tau^*$.

\subsection{Longitudinal correlator for time reversal by generalised Hahn echo}
We consider the case of starting from the ground state of $H(g_0=0)$. 
The echo in the transverse magnetisation is given by
\begin{align}
	\langle S_i^zS_{i+1}^z\rangle_{t_e}
	=\int_{-\pi}^\pi\frac{dk}{8\pi}e^{-ik}\left(\Sigma_x^k(t_e)+\im\Sigma_y^k(t_e)\right)
	=\int_{-\pi}^\pi\frac{dk}{8\pi}\left(\Sigma_x^k(t_e)\cos k+\Sigma_y^k(t_e)\sin k\right)\ .
\end{align}
The analysis of these integrals is mainly analogous to the previous section; however, for the
Hahn echo protocol the relevant saddle points are contributed by the integrals containing 
$\cos(2(\epsilon_k^g+\epsilon_k^{-g})\tau)$ without a shift of the echo time ($\nu=(t_e-\tau)/\tau=1$).
Since with $k_\pm^*=\pm\pi/2$
\begin{align}
	\left.\frac{d}{dk}\left(\epsilon_k^g+\epsilon_k^{-g}\right)\right|_{k=k_\pm^*}
	=\pm\left(\frac{1}{\epsilon_{k_\pm^*}^g}-\frac{1}{\epsilon_{k_\pm^*}^{-g}}\right)=0
\end{align}
and the corresponding test function 
$\sin k_{\pm}^*\cos\theta_{k_\pm^*}^{-g}\sin\phi_{k_\pm^*}^{g,g_0}(\cos\phi_{k_\pm}^*+1)/2\neq0$, the
stationary phase approximation yields
\begin{align}
	\langle S_i^zS_{i+1}^z\rangle_{t_e}-\langle S_i^zS_{i+1}^z\rangle_{\infty}
	\approx \tilde\beta^{g_0,g}\frac{\tilde\kappa(\tau)}{\tau^{1/2}}
\end{align}
with
\begin{align}
	\tilde\kappa(\tau)&\equiv\cos(2(\epsilon_{\pi/2}^g+\epsilon_{\pi_2}^{-g})\tau+\pi/4)\ ,\\
	\tilde\beta^{g}&\equiv\frac{\zeta^{g}}{4\sqrt{\pi}|\tilde\xi^{g}|^{1/2}}\ ,\\
	\tilde\zeta^{g}
	&\equiv\cos\theta_{\pi/2}^{-g}\frac{\cos\phi_{\pi/2}^{-g,g}+1}{2}\ ,\\
	\tilde\xi^{g}&\equiv\left.\frac{d^2}{dk^2}(\epsilon_k^g+\epsilon_k^{-g})\right|_{k=\pi/2}\ .
\end{align}
Note that in this case the frequency of the oscillatory term $\tilde\kappa(\tau)$ is given by the sum
of the spectra and for $g\gg1$ $\epsilon_{\pi/2}^g+\epsilon_{\pi_2}^{-g}\approx g$. Therefore, the echo
peak height at $t_e=2\tau$ oscillates with a high frequency, but the amplitude of the oscillations follows
a power law with exponent $-1/2$.
\end{document}